\documentclass{article}
\usepackage{spconf}
\usepackage{amsmath,amssymb,graphicx,url,times,booktabs,MnSymbol,tabularx}
\usepackage{acronym}
\usepackage{tikz,pgfplots,pgfplotstable}
\usepgfplotslibrary{groupplots}
\usetikzlibrary{patterns}
\pgfplotsset{compat=1.18} 
\usepackage[colorlinks=false,pdfborder={0 0 0}]{hyperref}
\usepackage{adjustbox}
\usepackage{subfig}
\usepgfplotslibrary{external} 
\newcommand{\BE}{\begin{equation*}\begin{aligned}}
\newcommand{\EE}{\end{aligned}\end{equation*}}

\acrodef{auc}[AUC-ROC]{area under the \acl{roc} curve}
\acrodef{aucpr}[AUC-PR]{area under the precision-recall curve}
\acrodef{asd}[ASD]{anomalous sound detection}
\acrodef{pauc}[pAUC]{partial area under the \acl{roc} curve}
\acrodef{roc}[ROC]{receiver operating characteristic}
\acrodef{sed}[SED]{sound event detection}
\acrodef{ev}[EV]{expected value}
\acrodef{fpr}[FPR]{false positive rate}
\acrodef{tpr}[TPR]{true positive rate}
\acrodef{pcc}[PCC]{Pearson correlation coefficient}

\title{F1-EV Score: Measuring the Likelihood of Estimating a Good Decision Threshold for Semi-Supervised Anomaly Detection}

 \name{Kevin Wilkinghoff $^{1}$ and Keisuke Imoto$^{2}$}
 \address{$^1$Fraunhofer FKIE, Fraunhoferstraße 20, 53343 Wachtberg, Germany\\
 $^2$Doshisha University, 1-3 Tatara Miyakodani, Kyotanabe, Kyoto, Japan\\
 kevin.wilkinghoff@ieee.org, keisuke.imoto@ieee.org}
\begin{document}

\ninept
\maketitle

\begin{sloppy}

\begin{abstract}
\Acl{asd} (\acs{asd}) systems are usually compared by using threshold-independent performance measures such as AUC-ROC.
However, for practical applications a decision threshold is needed to decide whether a given test sample is normal or anomalous.
Estimating such a threshold is highly non-trivial in a semi-supervised setting where only normal training samples are available.
In this work, F1-EV a novel threshold-independent performance measure for \acs{asd} systems that also includes the likelihood of estimating a good decision threshold is proposed and motivated using specific toy examples.
In experimental evaluations, multiple performance measures are evaluated for all systems submitted to the \acs{asd} task of the DCASE Challenge 2023.
It is shown that F1-EV is strongly correlated with AUC-ROC while having a significantly stronger correlation with the F1-score obtained with estimated and optimal decision thresholds than AUC-ROC.
\end{abstract}


\section{Introduction}
\label{sec:intro}
There are several performance measures for sound event detection \cite{mesaros2016metrics,ferroni2021improving} and \ac{asd} systems or binary classifiers in general \cite{koyejo2014consistent,paparrizos2022volume,boyd2013area,canbek2023ptopi}.
Usually, threshold-independent performance measures such as the \ac{auc} are used because they are more objective \cite{aggarwal2017outlier,ebbers2022threshold} compared to threshold-dependent measures such as the F1-score, which rely on a single chosen decision threshold.
However, it has been shown that the \ac{auc} has a few weaknesses \cite{lobo2008auc} as for example that \ac{auc} does not work well for imbalanced class distributions since equal weight is given for positive and negative samples \cite{lobo2008auc,saito2015precision}.
Moreover, for practical applications a decision threshold is needed to be able to decide whether a test sample is normal or anomalous.
In semi-supervised \ac{asd} settings, estimating this threshold boils down to separating the extreme values of the anomaly scores, e.g. scores larger than the 90th percentile, belonging to normal samples from the rest and hoping that the estimated decision threshold also works well for separating the scores of normal samples from the ones obtained with anomalous samples \cite{wilkinghoff2022choosing}.
Hence, it is implicitly assumed that the anomaly scores belonging to the normal training and test samples follow the same distribution.
Estimating a good decision threshold is a difficult task that is vital when developing a system for a practical \ac{asd} application.
If only \ac{auc} is used as a performance measure, this difficulty is not explicitly captured by the resulting score. 
\par
A na\"{i}ve solution to this problem is to use multiple performance measures, for example a threshold-independent and a threshold-dependent measure such as the \ac{auc} and the F1-score.
However, a single threshold-independent measure is much more favorable to objectively compare the performance of multiple systems.
In this work, we propose the F1-\ac{ev} score for measuring the performance of an \ac{asd} system.
Similar to \ac{auc}, F1-\ac{ev} is a threshold-independent performance measure but also takes the likelihood of estimating a good decision threshold into account.
\par
The contributions of this work are the following:
First, it is shown experimentally and through toy examples that \ac{auc} alone is not a sufficient performance measure for \ac{asd}.
Second, F1-\ac{ev}\footnote{An open-source implementation of F1-\acs{ev} is available at: \url{https://github.com/wilkinghoff/f1-ev}} a threshold-independent performance measure for anomaly detection is presented.
Furthermore, multiple performance measures are evaluated and compared using all systems submitted to the \ac{asd} task of the DCASE 2023 Challenge.
It is shown that bounding the F1-\ac{ev} score is important and F1-\ac{ev} is strongly correlated with \ac{auc} while also having a significantly higher correlation with the F1-score than \ac{auc}.
As a last contribution, fine-tuning the bounds of F1-\ac{ev} score is investigated in an ablation study.

\section{Performance measures}
\label{sec:definition}
Throughout this work, anomaly scores are positive scalar values, and their magnitude corresponds to the degree of anomaly.

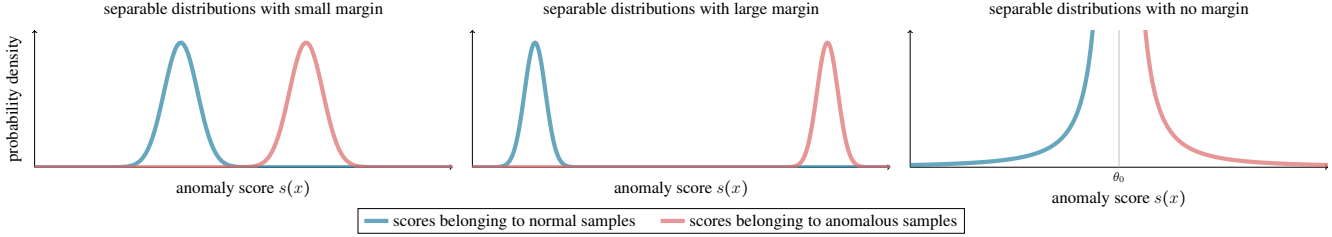
\begin{figure*}[!htb]
	\centering
    \begin{adjustbox}{width=\textwidth}
    \pgfmathdeclarefunction{gauss}{2}{%
  \pgfmathparse{1/(#2*sqrt(2*pi))*exp(-((x-#1)^2)/(2*#2^2))}%
}

\pgfmathdeclarefunction{xnorm}{0}{%
  \pgfmathparse{1/(pi-x)^2}%
}

\pgfmathdeclarefunction{xanom}{0}{%
  \pgfmathparse{1/(x-pi)^2}%
}

\begin{tikzpicture}
    \begin{groupplot}[
    xmin=-10,
    xmax=10,
    ymin=0,
    group style={
    group name=my plots,
    group size=3 by 1,
    xlabels at=edge bottom,
    ylabels at=edge left,
    horizontal sep=0.5cm,vertical sep=1cm,},
    ylabel={probability density},
    ylabel near ticks,
    height=5cm,
    width=12cm,
    xlabel={anomaly score $s(x)$},
    tick style={draw=none},
    yticklabel=\empty,
    xticklabel=\empty,
    label style = {font=\large},
    legend style = {font=\large},
    title style = {font=\large},
    axis line style={->},
    no markers,
	axis y line*=left,
    axis x line*=bottom,
    ]
    \nextgroupplot[title=separable distributions with small margin, legend style={at={(1.5,-0.3)},anchor=north,legend columns=2,/tikz/every even column/.append style={column sep=0.5cm}}]
    \addlegendentry{scores belonging to normal samples}
    \addlegendentry{scores belonging to anomalous samples}
    \addplot+[teal!85!blue!85!, opacity=0.7, domain=-10:10, samples=200,line width=3pt]{gauss(-3, 0.8)};
    \addplot+[teal!20!red!65!, opacity=0.7, domain=-10:10, samples=200,line width=3pt]{gauss(3, 0.8)};
    \nextgroupplot[title=separable distributions with large margin]
    \addplot+[teal!85!blue!85!, opacity=0.7, domain=-10:10, samples=500,line width=3pt]{gauss(-7, 0.5)};
    \addplot+[teal!20!red!65!, opacity=0.7, domain=-10:10, samples=500,line width=3pt]{gauss(7, 0.5)};
    \nextgroupplot[title=separable distributions with no margin, extra x ticks={pi},
    extra x tick style={%
        grid=major,width=3pt
        },
    extra x tick labels={$\theta_0$},
    xmin=0,
    xmax=2*pi,
    ymin=0,
    ymax=1/(pi-0.95*pi),]
    \addplot+[teal!85!blue!85!, opacity=0.7, domain=0:pi*0.95, samples=200,line width=3pt]{xnorm};
    \addplot+[teal!20!red!65!, opacity=0.7, domain=pi*1.05:2*pi, samples=200,line width=3pt]{xanom};
    \end{groupplot}
    \end{tikzpicture}
    \end{adjustbox}
    \caption{Three toy examples of perfectly separable anomaly score distributions, each with an \acs{auc} equal to 1. On the left, the margin between normal and anomalous scores is small. In the center, the margin between the distributions is large and thus estimating a good decision threshold is much easier. On the right, the only optimal decision threshold is $\theta=\theta_0$, which is a null set with measure zero, and thus estimating this threshold also has a likelihood of zero when assuming a continuous distribution with uncountable support for estimated thresholds. Note that when estimating a decision threshold, only a finite number of anomaly scores belonging to the distribution of normal samples are available and usually both distributions are more complex and overlap. This is the reason why estimating a good decision threshold is highly non-trivial.}
    \label{fig:anomaly_score_examples}
\end{figure*}

\subsection{\acs{auc}}
First, we will recall the definition of the \ac{auc} score.
For a set of $N\in\mathbb{N}$ threshold values $\theta(n)\in\mathbb{R}$ indexed by $n\in\lbrace1,...,N\rbrace$, let $x(n), y(n)\in[0,1]$ be the monotone increasing sequences of sorted \acp{fpr} and sorted \acp{tpr} of the \ac{roc}-curve resulting from the anomaly scores, respectively.
Define \BE\Delta x(n)&:=x(n+1)-x(n).\EE
More concretely, the sequence $(\theta(n))_{n=1,...,N}$ is set to all sorted anomaly score values belonging to the evaluation samples because these are the points where the intermediate statistics, i.e. \ac{tpr} and \ac{fpr}, change.
Compared to linearly scaled thresholds, which require an infinitesimal resolution to yield exact results, this particular choice of evaluation thresholds has two advantages \cite{ebbers2022threshold}: improved computational efficiency and improved accuracy of the \ac{roc} curve.
Then, using the trapezoidal rule, the \ac{auc} score can be approximated by calculating
\BE \text{AUC-ROC}(x,y)\approx\sum_{n=1}^{N-1}\frac{y(n+1)+y(n)}{2}\cdot\Delta x(n)\EE
as implemented in \cite{scikit-learn}.
\par
\ac{auc} does not require a specifically chosen decision threshold.
This allows for objective comparisons between different \ac{asd} systems but does not take into account the difficulty of estimating a good decision threshold.
The \ac{auc} score is equal to the probability that the anomaly score of a random normal sample is smaller than the anomaly score of a random anomalous sample  \cite{aggarwal2017outlier,hanley1982meaning}.
Hence, even when \ac{auc} equals $1$, the results can look very different because such a value only shows that there exists a threshold that perfectly separates the anomaly scores belonging to normal and anomalous samples as illustrated in \autoref{fig:anomaly_score_examples}.
This motivates the definition of other threshold-independent performance measures as proposed in the following two subsections.

\subsection{F1-\acs{ev}}
\begin{figure}[!t]
	\centering
    \begin{adjustbox}{width=\columnwidth}
    \begin{tikzpicture}
    \begin{axis}[
    xmin=1,
    xmax=20,
    ymin=0,
    ymax=1,
    ylabel={F1 score},
    ylabel near ticks,
    xlabel={decision threshold $\theta(n)$},
    xtick={1,3.3,5,7.8,11.4,13,17,19},
    xticklabels={$\theta(1)$,$\theta(2)$,$\theta(3)$,$\theta(4)$,$\theta(5)$,$\theta(6)$,$\theta(7)$,$\theta(8)$},
    axis line style={->},
    height=4cm,
    width=12cm,
	axis y line*=left,
    axis x line*=bottom]
	  \addplot[domain=1:3.3, gray!30, line width=0pt,samples=2,fill=blue,pattern=north east lines,pattern color=gray!30]{0.72727273}\closedcycle;
	  \addplot[domain=3.3:5,gray!30, line width=0pt,samples=2,fill=blue,pattern=north east lines,pattern color=gray!30]{0.8}\closedcycle;
	  \addplot[domain=5:7.8,gray!30, line width=0pt,samples=2,fill=blue,pattern=north east lines,pattern color=gray!30]{0.66666667}\closedcycle;
	  \addplot[domain=7.8:11.4,gray!30, line width=0pt,samples=2,fill=blue,pattern=north east lines,pattern color=gray!30]{0.75}\closedcycle;
	  \addplot[domain=11.4:13,gray!30, line width=0pt,samples=2,fill=blue,pattern=north east lines,pattern color=gray!30]{0.85714286}\closedcycle;
	  \addplot[domain=13:17,gray!30, line width=0pt,samples=2,fill=blue,pattern=north east lines,pattern color=gray!30]{0.66666667}\closedcycle;
	  \addplot[domain=17:19,gray!30, line width=0pt,samples=2,fill=blue,pattern=north east lines,pattern color=gray!30]{0.4}\closedcycle;
	  \addplot[domain=19:20,gray!30, line width=0pt,samples=2,fill=blue,pattern=north east lines,pattern color=gray!30]{0}\closedcycle;
	  \addplot[domain=1:3.3, teal!85!green!85!, line width=3pt,samples=2]{0.72727273};
	  \addplot[domain=3.3:5, teal!85!green!85!, line width=3pt,samples=2]{0.8};
	  \addplot[domain=5:7.8, teal!85!green!85!, line width=3pt,samples=2]{0.66666667};
	  \addplot[domain=7.8:11.4, teal!85!green!85!, line width=3pt,samples=2]{0.75};
	  \addplot[domain=11.4:13, teal!85!green!85!, line width=3pt,samples=2]{0.85714286};
	  \addplot[domain=13:17, teal!85!green!85!, line width=3pt,samples=2]{0.66666667};
	  \addplot[domain=17:19, teal!85!green!85!, line width=3pt,samples=2]{0.4};
	  \addplot[domain=19:20, teal!85!green!85!, line width=3pt,samples=2]{0};
    \end{axis}
\end{tikzpicture}
    \end{adjustbox}
    \caption{Example of computing F1-\ac{ev} using only very few decision thresholds for illustration purposes.}
    \label{fig:f1-ev}
\end{figure}
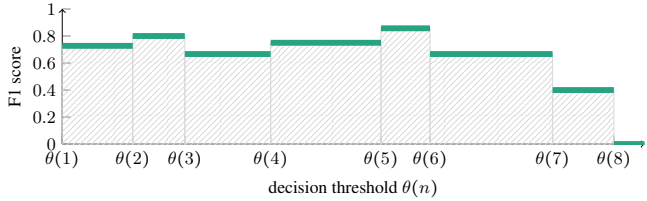
The idea of F1-\ac{ev} is to calculate the expected value of a random variable that models the F1-score of an anomaly detection system.
This is realized by calculating the F1-score for all possible threshold values $\theta(n)$ and computing the area under the resulting piece-wise constant F1-score function.
In \cite{davis2006relationship}, it has been shown that computing the \ac{auc} of the precision-recall curve by using linear interpolations between samples as done by the trapezoidal rule leads to over-optimistic results.
Since the F1-score is the harmonic mean of precision and recall, it thus makes more sense to also not use the trapezoidal rule.
We utilize a finite Riemann sum instead as illustrated in \autoref{fig:f1-ev}.
Since the empirical F1-score function is piece-wise constant with respect to the decision threshold, using more threshold values than the number of samples does not improve the accuracy of the Riemann sum.
\par
Let us now formally introduce F1-\ac{ev}.
Define the normalized distance between thresholds as \BE\Delta \theta(n):=\frac{\theta(n+1)-\theta(n)}{\theta(N)-\theta(1)}.\EE
Further, let $F1(\theta(n))$ denote the F1-score obtained when applying the decision threshold $\theta(n)$.
Then, the F1-\ac{ev} score is defined as
\BE\text{F1-EV}(\theta):=\sum_{n=1}^{N-1} \text{F}1(\theta(n))\cdot\Delta \theta(n).\EE
When assuming a uniform distribution for estimating a threshold in the interval $[\theta(1),\theta(N)]$, the value $\Delta \theta(n)$ corresponds to the likelihood of estimating the decision threshold $\theta(n)$.
In conclusion, F1-\ac{ev} is the expected value of a random variable that models the obtained F1-score of an anomaly detection system, as intended, and thus yields a score between $0$ and $1$ with higher values indicating better performance.

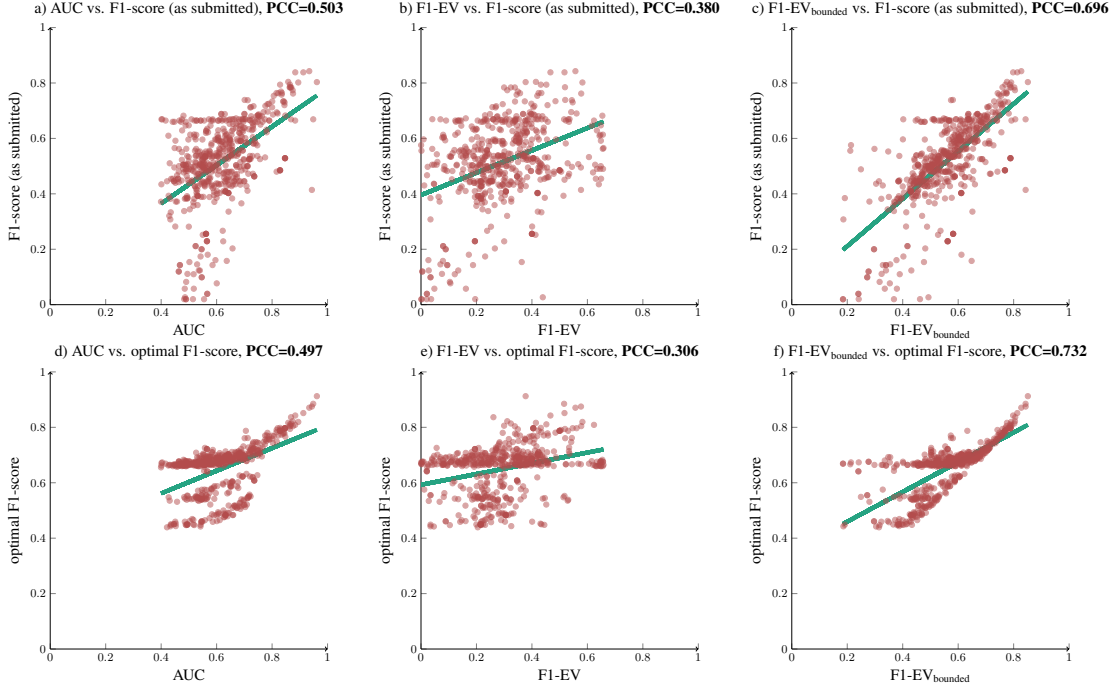
\begin{figure*}[!htb]
	\centering
    \begin{adjustbox}{width=0.83\textwidth}
    \begin{tikzpicture}
    \begin{groupplot}[
    xmin=0,
    xmax=1,
    ymin=0,
    ymax=1,
    group style={
    group name=my plots,
    group size=3 by 3,
    xlabels at=edge bottom,
    ylabels at=edge left,
    horizontal sep=2.5cm,vertical sep=1.8cm,},
    ylabel near ticks,
    axis line style={->},
    no markers,
    height=9cm,
    width=9cm,
	axis y line*=left,
    axis x line*=bottom,
    title style = {font=\large},
    label style = {font=\large},
    ]

    \nextgroupplot[title={a) AUC vs. F1-score (as submitted), \textbf{PCC=0.503}}, ylabel=F1-score (as submitted), xlabel=AUC]
    \addplot+[teal!85!green!85!,line width=3pt] table[row sep=newline,y={create col/linear regression={y=x2}}]{figures/auc-roc_vs_f1-sub.txt};
    \addplot+[only marks,mark=*,mark options={color=teal!35!red!85!,scale=1,opacity=0.5}] table[row sep=newline,x=x1,y=x2]{figures/auc-roc_vs_f1-sub.txt};

    \nextgroupplot[title={b) $\text{F1-EV}$ vs. F1-score (as submitted), \textbf{PCC=0.380}}, ylabel=F1-score (as submitted), xlabel=$\text{F1-EV}$]
    \addplot+[teal!85!green!85!,line width=3pt] table[row sep=newline,y={create col/linear regression={y=x2}}]{figures/f1-ev_vs_f1-sub.txt};
    \addplot+[only marks,mark=*,mark options={color=teal!35!red!85!,scale=1,opacity=0.5}] table[row sep=newline,x=x1,y=x2]{figures/f1-ev_vs_f1-sub.txt};

    \nextgroupplot[title={c) $\text{F1-EV}_\text{bounded}$ vs. F1-score (as submitted), \textbf{PCC=0.696}}, ylabel=F1-score (as submitted), xlabel=$\text{F1-EV}_\text{bounded}$]
    \addplot+[teal!85!green!85!,line width=3pt] table[row sep=newline,y={create col/linear regression={y=x2}}]{figures/f1-ev-bounded_vs_f1-sub.txt};
    \addplot+[only marks,mark=*,mark options={color=teal!35!red!85!,scale=1,opacity=0.5}] table[row sep=newline,x=x1,y=x2]{figures/f1-ev-bounded_vs_f1-sub.txt};

    \nextgroupplot[title={d) AUC vs. optimal F1-score, \textbf{PCC=0.497}}, ylabel=optimal F1-score, xlabel=AUC]
    \addplot+[teal!85!green!85!,line width=3pt] table[row sep=newline,y={create col/linear regression={y=x2}}]{figures/auc-roc_vs_f1-opt.txt};
    \addplot+[only marks,mark=*,mark options={color=teal!35!red!85!,scale=1,opacity=0.5}] table[row sep=newline,x=x1,y=x2]{figures/auc-roc_vs_f1-opt.txt};

    \nextgroupplot[title={e) $\text{F1-EV}$ vs. optimal F1-score, \textbf{PCC=0.306}}, ylabel=optimal F1-score, xlabel=$\text{F1-EV}$]
    \addplot+[teal!85!green!85!,line width=3pt] table[row sep=newline,y={create col/linear regression={y=x2}}]{figures/f1-ev_vs_f1-opt.txt};
    \addplot+[only marks,mark=*,mark options={color=teal!35!red!85!,scale=1,opacity=0.5}] table[row sep=newline,x=x1,y=x2]{figures/f1-ev_vs_f1-opt.txt};

    \nextgroupplot[title={f) $\text{F1-EV}_\text{bounded}$ vs. optimal F1-score, \textbf{PCC=0.732}}, ylabel=optimal F1-score, xlabel=$\text{F1-EV}_\text{bounded}$]
    \addplot+[teal!85!green!85!,line width=3pt] table[row sep=newline,y={create col/linear regression={y=x2}}]{figures/f1-ev-bounded_vs_f1-opt.txt};
    \addplot+[only marks,mark=*,mark options={color=teal!35!red!85!,scale=1,opacity=0.5}] table[row sep=newline,x=x1,y=x2]{figures/f1-ev-bounded_vs_f1-opt.txt};
    
    \end{groupplot}
    \end{tikzpicture}
    \end{adjustbox}
    \caption{Comparison of several different performance measures computed on the evaluation set of task 2 of the DCASE2023 Challenge. In the top row, threshold-independent performance measures are compared to the F1-score obtained with the submitted decision threshold. In the bottom row, threshold-independent performance measures are compared to the F1-score obtained with an optimal decision threshold.}
    \label{fig:experimental_results}
\end{figure*}
\subsection{Bounded F1-\acs{ev}}
When setting the range of possible thresholds, one has to define a smallest and a largest value of possible thresholds to compute a performance measure.
For F1-\ac{ev}, these values have been set to the smallest and largest anomaly scores of a given set of test samples.
However, many of these thresholds are very unlikely to be estimated.
One can expect that estimated decision thresholds lie in a certain range that depends on the estimation method.
Furthermore, if there are a few outliers in the anomaly score samples that are much smaller or much larger than all the other scores, this will skew the final performance measure drastically.
Therefore, we propose to define more robust boundaries for the interval of possible thresholds by only utilizing the anomaly scores belonging to the normal test samples.
Hence, the evaluation is independent of the particular set of normal training samples, which are assumed to follow the same distribution as the validation and test samples when estimating a decision threshold.
Furthermore, in \cite{wilkinghoff2022choosing} it has been shown that holding back a few normal training samples when training the system to obtain more realistic anomaly scores from these samples for estimating a decision threshold does not increase the resulting F1-score.
This indicates that the assumption made about the distributions is valid.
\par
We set the lower bound of the decision threshold, denoted by $\theta_\text{min}\in\mathbb{R}$, to be close to the sample mean $\mu$ of the anomaly scores belonging to the normal test samples.
Here we assume that, ideally, one would not estimate a decision threshold that is much smaller than this value.
An upper bound $\theta_\text{max}\in\mathbb{R}$ is more difficult to choose because the largest value of the anomaly scores may be an outlier and $+\infty$ is definitely too large.
Instead, we propose to utilize an upper bound close to the empirically optimal decision threshold $\theta_\text{opt}\in\mathbb{R}$, which is chosen as the center of the interval of possible thresholds yielding optimal results. 
Again, the underlying assumption is that, in an ideal world, one would not estimate a threshold much larger than this upper bound.
More concretely, we set
\BE \theta_\text{max}:=\theta_\text{opt}+\alpha\cdot\sigma\text{ and }\theta_\text{min}:=\mu-\alpha\cdot\sigma\EE
where $\sigma\in\mathbb{R}$ denotes the sample standard deviation belonging to the normal test samples and $\alpha\in\mathbb{R}$ is a hyperparameter to be set.
For the experiments conducted in this work, we manually set $\alpha=0.2$.
More details on choosing an appropriate value for $\alpha$ can be found in \autoref{subsec:ablation}.
Now, for $k=1,...,K$ with ${K=2+\lvert\lbrace n\in\lbrace1,...,N\rbrace:\theta_\text{min}<\theta(n)<\theta_\text{max}\rbrace\rvert}$ we set \BE\theta_\text{bounded}(k)=\begin{cases} \theta_\text{min}&\text{ if }k=1\\
\theta(k-1)&\text{ if }\theta_\text{min}<\theta(k-1)<\theta_\text{max}\\
\theta_\text{max}&\text{ if }k=K\end{cases}.\EE
The normalized distances between thresholds are adapted from the distances used for the basic F1-\ac{ev} score
\BE\Delta \theta_\text{bounded}(k):=\frac{\theta_\text{bounded}(k+1)-\theta_\text{bounded}(k)}{\theta_\text{bounded}(K)-\theta_\text{bounded}(1)}\EE
and the bounded F1-\ac{ev} score is set to
\BE\text{F1-EV}_\text{bounded}(\theta_\text{bounded}):=\sum_{k=1}^{K-1} \text{F}1(\theta_\text{bounded}(k))\cdot\Delta \theta_\text{bounded}(k).\EE
Again, this results in a score between $0$ and $1$ with higher values indicating better performance.


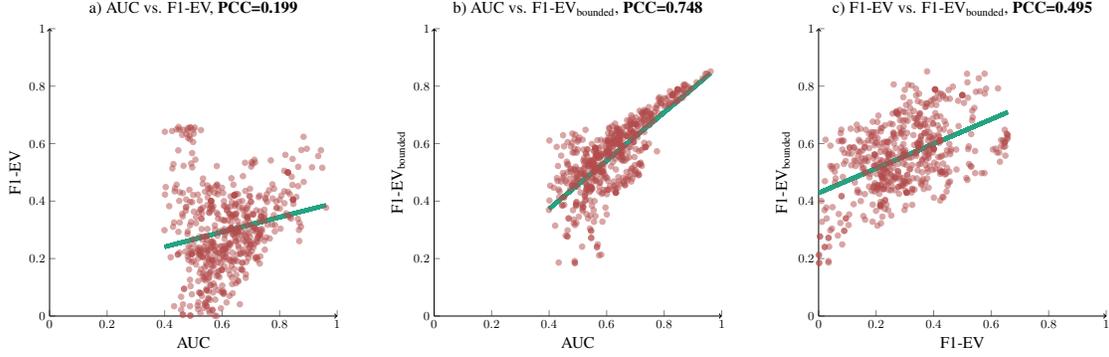
\begin{figure*}[!htb]
	\centering
    \begin{adjustbox}{width=0.83\textwidth}
    \begin{tikzpicture}
    \begin{groupplot}[
    xmin=0,
    xmax=1,
    ymin=0,
    ymax=1,
    group style={
    group name=my plots,
    group size=3 by 3,
    xlabels at=edge bottom,
    ylabels at=edge left,
    horizontal sep=2.5cm,vertical sep=1.8cm,},
    ylabel near ticks,
    axis line style={->},
    no markers,
    height=9cm,
    width=9cm,
	axis y line*=left,
    axis x line*=bottom,
    title style = {font=\large},
    label style = {font=\large},
    ]
    \nextgroupplot[title={a) AUC vs. F1-EV, \textbf{PCC=0.199}}, ylabel=F1-EV, xlabel=AUC]
    \addplot+[teal!85!green!85!,line width=3pt] table[row sep=newline,y={create col/linear regression={y=x2}}]{figures/auc-roc_vs_f1-ev.txt};
    \addplot+[only marks,mark=*,mark options={color=teal!35!red!85!,scale=1,opacity=0.5}] table[row sep=newline,x=x1,y=x2]{figures/auc-roc_vs_f1-ev.txt};

    \nextgroupplot[title={b) AUC vs. $\text{F1-EV}_\text{bounded}$, \textbf{PCC=0.748}}, ylabel=$\text{F1-EV}_\text{bounded}$, xlabel=AUC]
    \addplot+[teal!85!green!85!,line width=3pt] table[row sep=newline,y={create col/linear regression={y=x2}}]{figures/auc-roc_vs_f1-ev-bounded.txt};
    \addplot+[only marks,mark=*,mark options={color=teal!35!red!85!,scale=1,opacity=0.5}] table[row sep=newline,x=x1,y=x2]{figures/auc-roc_vs_f1-ev-bounded.txt};

    \nextgroupplot[title={c) $\text{F1-EV}$ vs. $\text{F1-EV}_\text{bounded}$, \textbf{PCC=0.495}}, ylabel=$\text{F1-EV}_\text{bounded}$, xlabel=$\text{F1-EV}$]
    \addplot+[teal!85!green!85!,line width=3pt] table[row sep=newline,y={create col/linear regression={y=x2}}]{figures/f1-ev_vs_f1-ev-bounded.txt};
    \addplot+[only marks,mark=*,mark options={color=teal!35!red!85!,scale=1,opacity=0.5}] table[row sep=newline,x=x1,y=x2]{figures/f1-ev_vs_f1-ev-bounded.txt};
    
    \end{groupplot}
    \end{tikzpicture}
    \end{adjustbox}
    \caption{Comparison of threshold-independent performance measures computed on the evaluation set of task 2 of the DCASE2023 Challenge.}
    \label{fig:experimental_results_independent}
\end{figure*}

\section{Experimental evaluations}
\label{sec:results}

\subsection{Experimental setup}

To compare different performance measures, we computed \acp{pcc} using all systems submitted to the \ac{asd} task of the DCASE2023 Challenge for machine condition monitoring \cite{harada2021toyadmos2,harada2023first,dohi2022mimii_dg,dohi2023description}.
This dataset consists of noisy audio recordings with a length between $6$ and $18$ seconds belonging to $14$ different machine types that are split into a development set and an evaluation set.
For each machine type contained in one of the sets, there is a source domain with $990$ normal training samples and a target domain with $10$ normal training samples that differs from the source domain by changing machine parameters or the background noise.
The test splits of the development and evaluation set consist of $100$ and $200$ samples for each machine type, respectively.
Any test sample is either normal or anomalous and belongs to the source or target domain.
The task is to predict whether a given test sample is normal or anomalous by using a single decision threshold for each machine type regardless of the domain the sample belongs to (domain generalization \cite{wang2021generalizing}).
To determine the ranking of the systems in the challenge, the harmonic mean of \ac{auc} and partial \ac{auc} \cite{mcclish1989analyzing} over all machine types was used.
Therefore, estimating a proper decision threshold to obtain a good F1-score is only optional and some participants abstained from doing this.
To not skew the results in this work, we only included the submissions with an F1-score greater than $0$ for the experimental evaluations.
Furthermore, we used the performance measures of all submitted systems for each machine type independently because another decision threshold is used for each machine type.

\subsection{Comparison of performance measures}

The experimental results are depicted in \autoref{fig:experimental_results} and \autoref{fig:experimental_results_independent}. The following observations can be made:
First, \ac{auc} has only a low to moderate correlation with the F1-scores belonging to the estimated decision thresholds (\ac{pcc}=0.503) and the optimal F1-scores (\ac{pcc}=0.497), which experimentally justifies the motivation for F1-\ac{ev} as a performance measure.
Second, the basic F1-\ac{ev} score has only very low correlation with \ac{auc} (\ac{pcc}=0.199) and also low correlation with the F1-scores (\ac{pcc}=0.380 and \ac{pcc}=0.306) showing that it is necessary to define proper bounds.
Third and most importantly, the bounded F1-\ac{ev} score, has a strong correlation with \ac{auc} (\ac{pcc}=0.748) and both F1-scores (\ac{pcc}=0.696 and \ac{pcc}=0.732).
Hence, F1-\ac{ev} better incorporates the difficulty of finding a good decision threshold than \ac{auc} and thus may also be a more suitable performance measure for \ac{asd}.
Note that some of the submissions may have estimated improper decision thresholds and thus the correlation between the bounded F1-\ac{ev} score and the F1-scores belonging to the estimated decision thresholds may be higher when only suitable estimation techniques are applied.
\par
The clusters shaped as horizontal lines in \autoref{fig:experimental_results} at an F1-score of approximately two-thirds, for example ranging from an \ac{auc} of $0.4$ to $0.9$ in Subfig. a), look odd but can be explained as follows:
It is possible that a chosen decision threshold works very well for the source domain but particularly bad for the target domain or vice versa.
This means that precision or recall is close to $1$ for both domains and the other value is close to $0$ for one domain and close to $1$ for the other domain, thus in total close to $0.5$ for both domains assuming that both domains have approximately the same number of samples.
Since the F1-score is the harmonic mean of precision and recall, this results in an F1-score of approximately two-thirds.

\subsection{Choosing the hyperparameter $\alpha$}
\label{subsec:ablation}

\begin{figure}[!t]
	\centering
    \begin{adjustbox}{width=\columnwidth}
    \begin{tikzpicture}
    \begin{axis}[
    xmin=0,
    xmax=2,
    ymin=0.4,
    ymax=1,
    ylabel={PCC with bounded F1-EV},
    ylabel near ticks,
    ymajorgrids,
    xlabel={$\alpha$},
    axis line style={->},
    height=5cm,
    width=12cm,
    no markers,
	axis y line*=left,
    axis x line*=bottom,
    legend style={at={(0.5,-0.3)},anchor=north,legend columns=3,/tikz/every even column/.append style={column sep=0.5cm}}]
      
    \addlegendentry{AUC}
    \addlegendentry{optimal F1-score}
    \addlegendentry{F1-score (as submitted)}
    \addplot+[teal!85!green!85!,opacity=0.7,line width=3pt] table[row sep=newline,x=x1,y=x2]{figures/alpha_vs_pcc_auc.txt};
    \addplot+[teal!20!red!65!,opacity=0.7,line width=3pt] table[row sep=newline,x=x1,y=x2]{figures/alpha_vs_pcc_f1-opt.txt};
    \addplot+[teal!85!blue!85!,opacity=0.7,line width=3pt] table[row sep=newline,x=x1,y=x2]{figures/alpha_vs_pcc_f1-sub.txt};
    \end{axis}
\end{tikzpicture}
    \end{adjustbox}
    \caption{Sensitivity of the bounded F1-\acs{ev} score with respect to $\alpha$.}
    \label{fig:alpha}
\end{figure}
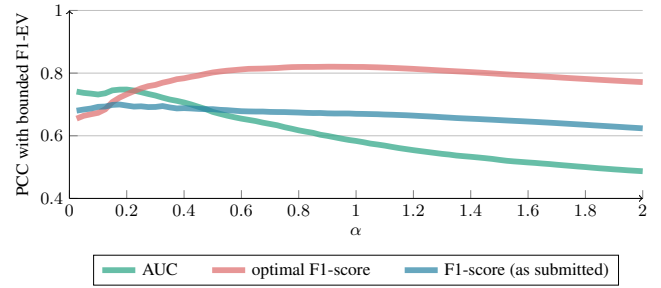
As an ablation study, the sensitivity of the bounded F1-\ac{ev} score with respect to the parameter $\alpha$ was investigated.
The results, depicted in \autoref{fig:alpha}, show that the \ac{pcc} between the bounded F1-\acs{ev} and \ac{auc} decreases for $\alpha>0.2$ showing that both measures are indeed different.
Furthermore, the \ac{pcc} between the bounded F1-\acs{ev} and the optimal F1-score increases for $\alpha<1$ and slightly decreases for $\alpha>1$ but keeps having a high correlation.
Lastly and most importantly, the \ac{pcc} between the bounded F1-\acs{ev} and the F1-score obtained with the submitted decision thresholds slightly increases for $\alpha<0.2$ and slightly decreases for $\alpha>0.2$ but to much less degree as it is the case for \ac{auc}.
In conclusion, the bounded F1-\acs{ev} is relatively stable with respect to $\alpha$ when considering estimated decision thresholds.
As the purpose of F1-\ac{ev} is to have a threshold-independent performance measure that is similar to \ac{auc} but also incorporates the difficulty of estimating a good decision threshold, we propose to use a relatively small value for $\alpha$, e.g. $\alpha=0.2$, as done in the other experiments of this work.

\section{Conclusion}
\label{sec:conclusion}
In this work, the threshold-independent performance measure F1-\ac{ev} for \ac{asd} systems that combines the advantages of both the threshold-independent \ac{auc} and threshold-dependent F1-score by correlating well with both of them was proposed.
In experiments conducted on the predictions of all systems submitted to the DCASE2023 \ac{asd} Challenge, it was shown that a bounded $\text{F1-EV}$ has a strong correlation with \ac{auc} while having a much higher correlation with the F1-scores based on estimated and optimal decision thresholds.
In conclusion, this performance measure has the potential to replace \ac{auc} as the de facto standard performance measure for \ac{asd}.
For future work, it is planned to conduct additional experiments on other datasets and further optimize the upper and lower bound of $\text{F1-EV}$.
In particular, choosing other distributions than a uniform distribution for estimating a decision threshold in the allowed range may be a promising direction.
Other experiments may focus on evaluating F1-\acs{ev} for settings with a strong imbalance between normal and anomalous samples and investigate specific machine types or domains individually.

\section{Acknowledgments}
The authors would like to thank Fabian Fritz and Frank Kurth for their valuable feedback.

\bibliographystyle{IEEEbib}
\bibliography{refs}

\end{sloppy}
\end{document}